# TIME SERIES DATA MINING FOR THE GAIA VARIABILITY ANALYSIS

*Krzysztof Nienartowicz, Diego Ordóñez Blanco, Leanne Guy, Berry Holl, Isabelle Lecoeur-Taïbi, Nami Mowlavi, Lorenzo Rimoldini, Idoia Ruiz, Maria Süveges, Laurent Eyer*

Department of Astronomy, University of Geneva, CH-1290 Versoix, Switzerland

## ABSTRACT

*Gaia is an ESA cornerstone mission, which was successfully launched December 2013 and commenced operations in July 2014. Within the Gaia Data Processing and Analysis consortium, Coordination Unit 7 (CU7) is responsible for the variability analysis of over a billion celestial sources and nearly 4 billion associated time series (photometric, spectrophotometric, and spectroscopic), encoding information in over 800 billion observations during the 5 years of the mission, resulting in a petabyte scale analytical problem. In this article, we briefly describe the solutions we developed to address the challenges of time series variability analysis: from the structure for a distributed data-oriented scientific collaboration to architectural choices and specific components used. Our approach is based on Open Source components with a distributed, partitioned database as the core to handle incrementally: ingestion, distributed processing, analysis, results and export in a constrained time window.*

## 1. INTRODUCTION

Gaia's Data Processing and Analysis Consortium (DPAC) provides a certain level of unification for data exchange between its six Data Processing Centers (DPCs) via a common data model shared among the Data Processing Centers. Every DPC is a producer of and a subscriber to relevant data within the consortium, which is received and re-distributed by the central repository managed by DPC ESAC (DPCE), located in Madrid. On the other hand, nine Coordination Units are governing the scientific analysis of the data via tools that the DPCs should provide, concentrating on implementation of scientific methods and algorithms. Clearly, there is a strong link between the science that can be produced and capabilities of the software it runs on. The scientific analysis done by CU7 is supported by the DPC Geneva (DPCG). CU7/DPCG try to leverage their close ties to boost the scientific capabilities via an unorthodox mixture of technologies. .

The idea is to have a self-describing, iterative, repeatable and mostly-automated system, conceptually similar to the one of the Fourth Paradigm postulated by Jim Gray [1]. Following M. Stonebreaker's Big Data characteristics, we put our foot in each of the Big Data *Three V* [2] fields:

- *Variety*: we deal with many data sources, data formats, plethora of data types, several data analysis languages (understood as general data diversity here);
- *Velocity*: we have a limited time window for data transformation during ingestion, process several iterations, then analyze and validate the results, and transform the results during export using different data access patterns;
- *Volume*: with both input and output aggregated, the sheer size of the data produced oscillates between 300TB to 1PB consisting of hundreds of billions of entities. We also have to deal with the *unknown* as this range depends on many, yet to be discovered, scientific factors.

## 2. THE VARIABILITY ANALYSIS

### 2.1. Data model

Astrophysical *Sources* of light that belong to a specific survey are grouped in *Catalogs*. Each source has an *owning catalog* that defines its identity but can also be referenced in any *derived catalogs* that are projections of sources with common features. Each source might have a collection of time series. Catalogs define metadata dictionaries that describe sources within them. We can have hierarchies of catalogs, so that sources are grouped naturally, i.e. by their types, usually with decreasing size due to specialization in subcatalogs. A Catalog provides a natural notion of input dataset for us.

Catalogs with metadata, sources and time series are input to *Runs*. A run is an execution of algorithms with a specific configuration on a chosen *Input catalog*. Certain runs might be designated as *Parent Runs, containing* results from which subsequent runs inherit. Additionally, runs can produce new catalogs either because of their *Selection criteria* or simply because they are data *Import runs*. Results stored in a run are, to some extent, symmetric with the input: we have *source results* with associated *time series results* and a plethora of other dependent results. Runs also hold aggregated information about all results within them as well as system information.
Figure 1 shows a simplified view of the data model described here.

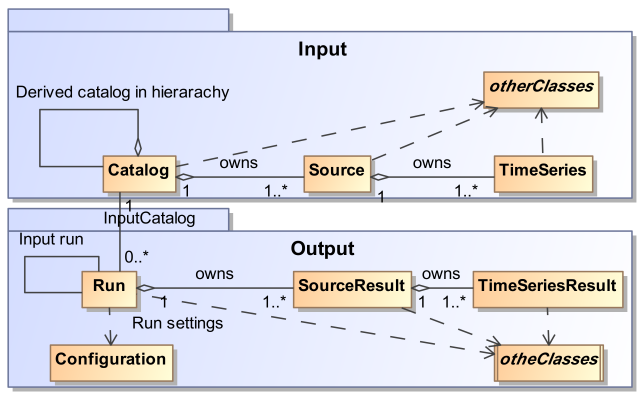

**Figure 1: Simplified data model**

### 2.2. Data distribution and partitioning

The data storage platform of choice is PostgreSQL. We use a Symmetric Scatter-Gather Enterprise Integration Pattern [3] for many processing aspects using Apache ActiveMQ as the queuing middleware and Apache Camel as an abstraction layer to create dataflows. Despite the excellent feature-set and extensibility of PostgreSQL, neither distribution nor parallel query execution is natively supported, contrary to some commercial databases and NoSql solutions. We evaluated several solutions and the most promising turned out to be a parallel, shared-nothing fork of PostgreSQL: Postgres-XC and its fork Postgres-XL that could be treated as NewSQL approaches [2]. We have successfully run basic tests with multimillion-source catalogs on both platforms, even though both are in relatively early stages of development and still lack robustness.

Tables in both systems can be replicated or distributed. Distribution in our case is based on a hash value of the Gaia source identity, which is largely based on a nested sky indexing Healpix [4] – all voluminous entities have the source Id as part of their identity, other lower cardinality entities like Catalogs or Runs are replicated over all nodes of the database cluster.

Distribution in this case also means parallelization of certain queries as well as load balancing of the ingestion of data.

Partitioning allows better vertical scaling of database nodes. Sources, source results, dependent entities and import entities are equi-width partitioned based star density estimates over the whole sky and the Gaia scanning law. Certain entities must be partitioned with additional overflow partitions to deal with the uncertainty of their positions. This approach is similar to chunk overlap used e.g., in SciDB [5] to avoid joins over additional, changing sets of partitions. Scientific analysis from the CU7 perspective can be divided into two main types described in the following two sections:

### 2.3. Source-by-source batch processing

Each light source is analyzed independently of all other sources. This happens in batches with preconfigured settings identical for all sources. Sources in batches are correlated only to optimize the IO path to fetch them and store results. Obviously, each source can yield a different analysis path depending on its characteristics. The results will be sampled and visualized, as an astronomer's eye is still a good validation tool. Ultimately, each source classified as one of the few dozen known variable classes [6] or identified as an outlier. In the latter case further analyses are performed to determine the nature of these sources, possibly identifying some new (sub-)classes.

The selection of high-level tasks of the variability analysis is shown in Figure 2. The initial step is to identify an object as being variable based on several statistical tests. Once variable candidates are identified, their variability is characterized by various statistical and model parameters. Simple models such as a trend or more complex models such as multi-harmonic periodic Fourier models are tested to determine whether they can sufficiently describe the observed variability. This characterization task produces a set of parameters, called attributes that are used in the next step of the processing: classification. The classification algorithms compute membership probabilities for each object on the basis of the specific values of its attributes. Supervised methods and associated training tools have already been implemented, whereas unsupervised and semi-supervised methods are under development, including time-series motif mining. At each step a global view on the data might be needed to re-asses results obtained during processing.
For more details of the tasks we refer to Eyer et al. 2012 [7].

### 2.4. Global Variability Studies

Global analysis both precedes and wraps around source-by-source processing, as might be conducted at any stage of the dataflow: during import, processing or data export. Results of global data mining are stored and can be visualized independently regardless how they were obtained.
We have several methods in Global Variability studies: *Validation*, *Monitoring* and *Posteriori* analysis.

*2.4.1. Validation*
To prevent scientific or system regressions we validate new generations of data using a *Golden Catalog* and *Golden Run*

comprising relatively small well-defined datasets ($< 10^6$ sources) representing either an established scientific benchmark or carefully crafted CU7 benchmarks. We run distributed validation against such datasets, comparing individual source results as well as the shape of the global distributions.

*2.4.2. Monitoring*
Of several hundred attributes we have derived a set of metrics that are crucial for the first assessment of the input data and source-by-source processing results. Mining of such metrics happens in *real-time* through *Monitoring*, which subscribes to the partial batch aggregates published by the distributed framework and finally merges them. We use the same framework to publish and aggregate system performance metrics.

*2.4.3. Posteriori analytics*
All ad-hoc queries on the input data, processing results or already existing aggregates from either monitoring or previous mining steps fall into this category.
A mix of methods is utilized:
- SQL queries which use PostgreSQL extensions such as embedded Java functions (plJava)
- SQL queries embedded in R
- GNU R stored procedures or R functions called from within SQL (plR).
- Selection of PostgreSQL extensions for spatial indexing as well as KNN-searches with various metrics developed by us, including time series SAX [8].

## 3. THE VARIABILITY PROCESSING DETAILS

Variability processing at CU7/DPCG is unique within the Gaia consortium due to the requirement of operating on the luminosity, spectra and radial velocity time series and other entities derived from these. The majority of the challenges we face revolve around time series and sources. Selected steps of the variability processing are listed below.

### 3.1. Data import

Algorithm verification and validation require both simulated and existing survey data. Such a heterogeneity of data triggered an effort to create an *Extract, Transform, Load* (ETL) distributed framework to map incoming data to the structure of our database, based on Enterprise Integration Patterns with Apache Camel and a number of supported components. De-localization of data for data curation is achieved using queuing and a Sun Grid Engine to access shared computing resources during import and export.

The Gaia data we receive consists of compressed files of serialized Java objects. The layout of these 1500 (and growing number of) types has evolved and will continue to evolve with the software and scientific discoveries. The *DataExchange* ETL system is capable of loading data into our database independently of changes in the data structures.

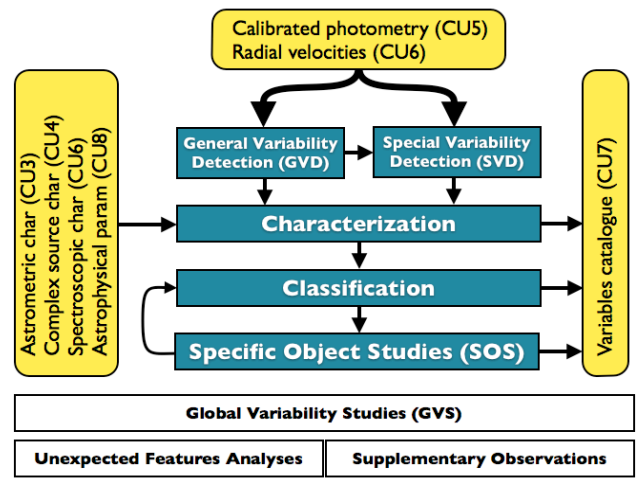

Figure 2: Tasks of the source variability analysis

### 3.2. Reconstruction of time series

All CU7/DPCG processing revolves around time series. Time series are, by definition, observations that are ordered in time. The peculiarity of the Gaia scanning law [9] results in streams of observations, (called *transits)* as they originate from light transiting the CCD detectors on board the satellite ordered in time, but not grouped by the sources of light from which they originate.
All the transits of each source must be thus gathered, ordered by time, compressed and stored. This *reconstruction* is a very significant task, dealing with more than 800 billion events with information scattered over several entities, to reorganize and cross-match around 4 billion time series. Additionally, Solar System barycentric light-travel time corrections are applied so that we can use standard time series analysis algorithms. The reconstruction is achieved with advanced SQL in parallel over a number of partitions.

### 3.3. Parallel computation and Global Variability studies

In addition to the standard HPC/batch processing with a job granularity of 100-1000 sources over few hundred cores, processing results are stored and analyzed a posteriori. All information about the environment, algorithm configuration, software versions and dependencies is stored together with the results to allow deep correlation analysis of software versions and results.

Analytics are conducted using SQL, either embedded in R, Java or extended with R, PL/Java and spatial and other non-standard indexing schemes, thanks to PostgresSQL extensibility.

Global variability studies are based on a comparative approach of global statistical metrics obtained via methods presented here.

### 3.4. Quality assurance, Scientific monitoring, Visualization

Various metrics can be obtained in *real-time* by streaming partial-processing results from processing nodes to pre-configured aggregating components, similar to a *Complex Events Processing* systems approach. Explicit *scatter-gather* logic was implemented to steer the contents of the samples over which we aggregate. Visualization tools based on R or Google Web Toolkit fetch either already aggregated data or store results in unified structures for aggregation. We developed support for scalars, 1D and 2D histograms, orthogonal grids and classification confusion matrices.

### 3.5. Data export

Our pipeline stores the results in a local data store. At the end of each processing cycle, following refinement and verification, they are sent back to ESAC to populate the Gaia catalog. As for the input data, the output data must also follow the project-wide agreed format. Data is retrieved, transformed to the corresponding type (we export almost a hundred different types), serialized and compressed before being transmitted to ESAC, from where it will ship to any subscribing DPC. Efficient use of configuration, parallelization and distribution techniques is required for timely export of the results.

### 3.6 Collaboration tools

Cooked-for-development subsets of data are distributed as ready-to-use PostgreSQL *virtual appliances*. Scientists can just deploy and run a scaled-down version of the system on laptops in the same way we run the system at DPCG.

### 3.7 Hardware platform

Initially we deployed our data platform on four nodes, with 40 hypethreaded cores per node, 256GB, with Infiniband interconnect with local storage and independent SAS controllers for Write Ahead Log and SSDs.
We use two Dell MD3260 disk enclosures with eight ports each, 60 3TB SATA disks connected to nodes in pairs. For High Availability (HA) we introduced primary-backup database instance setup conceptually similar to RAID-Z [10] where each node plays the role of a primary database and of a secondary (backup) for a node connected to a different enclosure, depicted on Figure 3.
For this architecture we conducted an extensive set of low-level IO tests to determine the optimal configuration of both the OS settings as well as those of the MD3260. This type of enclosure is not ideal due to its shared-disk nature but might prove advantageous for fabric unification.

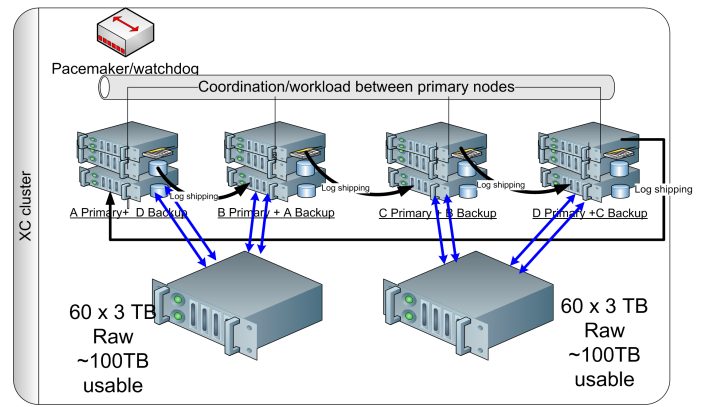

**Figure 3: Deployment for High Availability**

Furthermore we noticed a variation in IO performance for different cabling/firmware/filesystem/OS settings of up to 300%.

### 4. CONCLUSIONS

This paper presented our frame of the Big Data processing in the context of photometric time-series mining for the Gaia mission. Spectrum of challenges this variability study exposes has been addressed in technical and scientific terms. This has been done by applying a mixture of good design practices and a profound understanding of both strong and weak points of the off-the-shelf public domain software. We believe we can reach the level postulated by The Fourth Paradigm to have an open and fully repeatable processing chain for distributed batch processing and analytics.